\documentclass[aps,twocolumn,notitlepage,pra,superscriptaddress,footinbib]{revtex4-1}
\usepackage{amsmath,amssymb,bm}
\usepackage{graphicx}
\usepackage{epstopdf}
\usepackage{latexsym}
\usepackage[normalem]{ulem} 
\usepackage[usenames,dvipsnames]{color}
\usepackage[colorlinks=true,citecolor=black,
urlcolor=black,linkcolor=black]{hyperref}
\usepackage{natbib}
\usepackage{braket}
\usepackage{comment}
\usepackage{wrapfig}
\usepackage{braket}
\begin{document}

\newcommand{\Rev}[1]{{\color{blue}{#1}\normalcolor}} 
\newcommand{\Com}[1]{{\color{red}{#1}\normalcolor}} 
\title{Enhancing Initial-State Sensitivity through Time-Dependent Hamiltonian Readout in Ising Spin Chains}
\author{H. V. Do}
\affiliation{Department of Physics, University of Massachusetts, Boston, Massachusetts 02125, USA}
\date{June 26, 2026}

\begin{abstract}    
    Local observables can lose sensitivity to an initial state during strongly interacting many-body evolution even though the global dynamics remain unitary. We show that this sensitivity can be enhanced through a time-dependent Hamiltonian readout. Two orthogonal product states are first evolved under a slanted-field Ising Hamiltonian, where their distinction becomes strongly suppressed as observed through several local observables, including subsystem magnetizations and correlation functions, and are then quenched to the transverse-field Ising model at a tunable time. Exact simulations of chains up to $N=12$ show that the optimized time-averaged separation after the switch exceeds the residual slanted-field baseline for every observable and system size tested. In the strongest channels, the standardized readout separation remains robust over the accessible size range, with no clear systematic suppression at larger $N$. The enhancement recurs in widely separated late-time windows and persists qualitatively for open boundaries. These results establish Hamiltonian switching as an observable-selective mechanism for enhancing initial-state sensitivity without time reversal or implying recovery of the full reduced state.
\end{abstract}

\maketitle

\section{Introduction}

A common challenge in quantum experiments is understanding why information initially accessible in a quantum system becomes difficult to observe. In some cases, the information is irreversibly degraded by noise or decoherence; in others, it remains coherently stored but becomes delocalized across many degrees of freedom and hidden from local measurements. In realistic settings, the system of interest is rarely perfectly isolated, while the full joint state of the system and its environment is generally inaccessible to measurement. Local measurements on the system of interest may therefore lose sensitivity to the initial state even when some of the information remains coherently stored in correlations with its surroundings. This distinction underlies recent protocols for distinguishing coherent information delocalization from decoherence \cite{Yoshida2019,Alonso2019,Landsman2019,Yan2020}. It motivates an operational question: if information has not irreversibly leaked into an uncontrolled external environment, but instead has coherently delocalized into a larger and partially controllable system, can controlled changes of global parameters make part of that information locally accessible again?

One established route to diagnosing coherently delocalized information is to use out-of-time-ordered correlators, Loschmidt echoes and related time-reversal protocols \cite{Yoshida2019,Yan2020,Yan2019,Swingle2016,Xu2024,Garttner2018,LewisSwan2020,Dominguez2021,Geier2024}. These approaches have provided powerful probes of many-body information spreading, but they often require either an approximate inversion of the prior dynamics, an echo sequence tailored to the system, or additional operations that may not be available in a given experimental platform. In many settings, the detailed delocalizing dynamics are not fully known, and the available controls are restricted to global parameters such as fields, detunings, or coupling strengths, while measurements remain limited to local observables, snapshots, and few-body correlations. This motivates a complementary question: can switching experimentally accessible control parameters serve as a readout operation, not by undoing the prior dynamics, but by mapping part of the information hidden in many-body correlations back into local observables?

In this work, we study a closed spin chain as a controlled system--bath analogue. A chosen local subsystem plays the role of the accessible system, while the remaining spins act as an internal bath. The full chain evolves unitarily, so information about the initial state is not destroyed \cite{Deutsch1994,Braunstein2006}. However, this global preservation does not imply that the information remains accessible to local measurements. As interactions generate entanglement across the spin chain, information about the initial state can become delocalized over many degrees of freedom, causing local observables and few-body correlators to lose sensitivity to the initial state \cite{Mori2018,Eis14,Kaufman2016}. From the perspective of the chosen subsystem, this mirrors realistic settings where the full system-plus-environment state is generally inaccessible \cite{Kaufman2016,Popescu2006}. Because the full spin chain remains closed and controllable in the model, any loss of local sensitivity can be attributed to coherent redistribution of information within the chain rather than to uncontrolled environmental loss. This closed-system setting therefore allows us to separate coherent delocalization into controllable degrees of freedom from irreversible leakage of information into an uncontrolled environment.

Here we study this question numerically in finite-size, time-dependent Ising spin chains motivated by programmable Rydberg atom arrays \cite{Browaeys2020,Browaeys2016,Eba20,Scholl2021,Slagle2021,Blu21,Monika2020,Zhang2024}, where laser parameters can tune the dynamics between regimes with rapid entanglement growth and regimes with more structured evolution. We focus on a Hamiltonian-switching protocol in which an initial slanted-field Ising evolution delocalizes the initial-state dependence into many-body degrees of freedom and suppresses its visibility in selected local and two-body observables, followed by a quench to a transverse-field regime used as a readout Hamiltonian. The protocol is not a time-reversal protocol and does not attempt to undo the many-body evolution. Instead, it asks whether a controlled change of Hamiltonian parameters, such as the applied global fields, can partially restore the sensitivity of experimentally accessible observables to the initial state.

Across the accessible system sizes, the transverse-field readout substantially enhances the sensitivity of selected local and two-body observables to the initial state relative to the slanted-field baseline. In the strongest channels, the two post-quench mean signals
are separated by several pooled dynamical fluctuation widths. For $M_x^A$ at $N=8$, the dynamical bands associated with the two distinct initial states are nearly nonoverlapping, while clear standardized
separation persists through $N=12$. The Hamiltonian switch therefore opens a robust and controllable readout channel for initial-state information that had become nearly inaccessible in the preceding observable dynamics. In selected channels, this conversion enables
nearly complete discrimination of the two initial states from the measured dynamical signal.

More broadly, these results motivate the study of time-dependent Hamiltonian sequences as tools for probing information accessibility \cite{Scholl2021}. Instead of using a fixed Hamiltonian only to generate many-body evolution, one can use controlled parameter changes as active readout operations: different parameter paths may map different components of delocalized many-body information back into local or few-body observables. This perspective may be useful whenever apparent information leakage is not fully irreversible, but instead corresponds to information becoming hidden in a contained and partially controllable environment. In such cases, the goal is not to replace quantum error correction or to recover information lost to an uncontrolled bath, but to diagnose whether information remains coherently stored within accessible degrees of freedom and whether tailored dynamics can make part of it locally measurable again. Thus, Hamiltonian-switching readout provides a finite-size diagnostic of coherent many-body memory, can help distinguish different initial states whose local observable dynamics have become nearly indistinguishable, and motivates both theoretical searches for optimized readout sequences and the experimental engineering of precise time-dependent fields, detunings, and interactions in programmable atom arrays.

More broadly, the question of how initially accessible information can become hidden by scrambling while remaining recoverable under unitary dynamics also underlies black-hole information-retrieval thought experiments and teleportation-inspired decoding protocols \cite{HaydenPreskill2007,YoshidaKitaev2017}. Although the present protocol is not a model of gravitational dynamics, it provides a controlled many-body setting in which the accessibility of hidden information can be actively modified through a simple change of Hamiltonian.

\section{Rydberg-Ising Model and Hamiltonian-switching protocol}\label{sec:model}

\subsection{Ising field model} 

We consider a finite one-dimensional array of $N$ effective two-level systems governed by an Ising Hamiltonian with tunable transverse and longitudinal fields,
\begin{equation}
H(B_x,B_z)
=
J\sum_{i=1}^{N_b}\sigma_i^z\sigma_{i+1}^z
+
B_x\sum_{i=1}^{N}\sigma_i^x
+
B_z\sum_{i=1}^{N}\sigma_i^z .
\label{eq:ising-general}
\end{equation}
Here \(J\) is the nearest-neighbor Ising interaction strength, \(B_x\) and \(B_z\) are global transverse and longitudinal fields, and \(\sigma_i^\alpha\) are Pauli operators acting on site \(i\). Throughout the numerical results, time is reported in units of $1/J$, and we set $J=1$. Unless otherwise stated, all main-text results use a ring geometry with periodic boundary conditions, for which \(N_b=N\) and \(\sigma_{N+1}^{\alpha}\equiv\sigma_1^{\alpha}\). In this geometry, every spin has two nearest neighbors, and the interaction-induced longitudinal field generated by the Rydberg interaction is spatially uniform. 

As a boundary-condition check, we also perform representative open-chain calculations, where \(N_b=N-1\) and the two edge spins have only one nearest-neighbor interaction. These calculations are discussed only in Appendix~\ref{app:open-chain} and are used to verify that the qualitative readout behavior is not a boundary artifact.

The two field components play distinct roles in the dynamics. By tuning \(B_x\) and \(B_z\), the model can be moved between regimes with different entanglement growth and different degrees of local memory in experimentally accessible observables.

The first is the slanted-field Ising model (SFIM),
\begin{equation}
H_{\rm SFIM} = H(B_x=1/\sqrt{2},B_z=1/\sqrt{2}),
\label{eq:sfim}
\end{equation}
where both transverse and longitudinal fields are of equal strength and the total field is comparable to the interaction scale. Prior studies have shown that, in this regime, eigenstates near the middle of the slanted-field Ising spectrum exhibit large bipartite half-chain entanglement entropy \cite{SFIM1,SFIM2,SFIM3} consistent with volume-law behavior \cite{Mitra2025}. For initial product states with broad spectral support, the slanted-field regime therefore provides a useful setting for rapid entanglement generation and for delocalizing initial-state dependence into many-body degrees of freedom.

The second is the transverse-field Ising model (TFIM)\cite{TFIM0,TFIM01},
\begin{equation}
H_{\rm TFIM}
=H(B_x =1 ,B_z = 0),
\label{eq:tfim}
\end{equation}

obtained by setting the longitudinal field to zero. Compared with the slanted-field regime, the transverse-field limit is symmetric, integrable and has more structured dynamics, leading to more constrained entanglement growth for the finite systems studied here \cite{Mitra2025,TFIM1,TFIM2}. We use this regime not as a time-reversal operation, but as a readout Hamiltonian: after an initial evolution under \(H_{\rm SFIM}\), the system is quenched to \(H_{\rm TFIM}\) to test whether selected local and two-body observables regain partial sensitivity to the initial state. Unlike time-reversal or echo protocols, which require detailed knowledge and control sufficient to approximately invert the prior dynamics, the protocol considered here only changes a global Hamiltonian parameter. In this sense, the readout is not designed to undo the evolution, but to test whether a simple experimentally accessible Hamiltonian change can make part of the delocalized initial-state dependence visible again.

The distinction between these two regimes \cite{Alba2019} is central to the protocol. The slanted-field evolution acts as the stage in which initial-state dependence becomes delocalized into many-body degrees of freedom and becomes weakly visible in chosen observables. The transverse-field evolution is then used as a controlled Hamiltonian switch that can map part of this hidden information back into experimentally accessible local channels. Throughout the paper, we refer to this effect as partial readout or partial re-accessibility, rather than information recovery in the thermodynamic or time-reversal sense.

\subsection{Initial states} 

We probe local accessibility by comparing the evolution of two initial product states polarized along opposite \(y\) directions,
\begin{equation}
|\psi_+(0)\rangle = |+y\rangle^{\otimes N},
\qquad
|\psi_-(0)\rangle = |-y\rangle^{\otimes N},
\label{eq:initial-states}
\end{equation}
where
\begin{equation}
|\pm y\rangle = \frac{1}{\sqrt{2}}
\left(|0\rangle \pm i |1\rangle\right).
\end{equation}
These states are useful for three reasons. First, they are orthogonal many-body states, so they are initially perfectly distinguishable at the global level. Second, they are experimentally simple product states that can be prepared. Third, they have broad support over the eigenstates of $H_{\rm SFIM}$. This broad spectral support makes them effective probes of entanglement generation and the delocalization of initial-state dependence during the slanted-field evolution.

\subsection{Hamiltonian-switching protocol}

\begin{figure}[h]
    \centering
    \includegraphics[width=\linewidth]{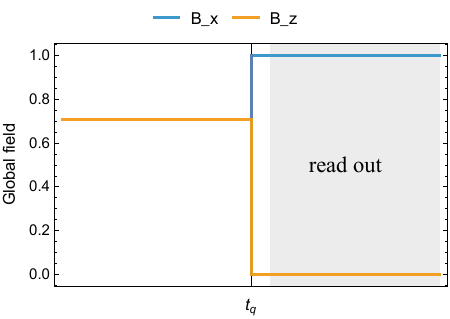}
    \caption{
    Time-domain schematic of the Hamiltonian-switching protocol. 
    From \(t=0\) to the quench time \(t_q\), the system evolves under the SFIM Hamiltonian with \(B_x=B_z=1/\sqrt{2}\). 
    At \(t_q\), the fields are switched to the TFIM regime, \(B_x=1\) and \(B_z=0\), while the interaction scale \(J\) is kept fixed at $J=1$. 
    The subsequent TFIM evolution is used as a readout stage.
    }

    \label{fig:quench-protocol}
\end{figure}

The protocol consists of two stages (See Fig. \ref{fig:quench-protocol}). In the first stage, each initial state evolves under the slanted-field Hamiltonian for a time \(t_q\),
\begin{equation}
|\psi_\pm(t_q)\rangle = e^{-iH_{\rm SFIM}t_q} |\psi_\pm(0)\rangle
\label{eq:sfim-evolution}
\end{equation}
Here \(t_q\) is the quench time, or equivalently the duration of the initial delocalizing evolution.

In the second stage, the Hamiltonian is switched from \(H_{\rm SFIM}\) to \(H_{\rm TFIM}\). 
For a time \(\tau\) after the quench, the state evolves as
\begin{equation}
|\psi_\pm(t_q+\tau)\rangle = e^{-iH_{\rm TFIM}\tau} |\psi_\pm(t_q)\rangle.
\label{eq:tfim-readout}
\end{equation}
This second evolution is used as a readout operation. It is not intended to reverse the earlier delocalizing dynamics; rather, it tests whether a simple global change in Hamiltonian parameters can partially restore the sensitivity of selected local and two-body observables to the initial state. To avoid relying on the immediate transient following the quench, we allow a post-quench waiting time $t_{\rm wait}$ before evaluating the readout signal. Observables are then averaged over a window of width $\Delta$,
\begin{equation}
t\in \left[ t_q+t_{\rm wait}, \, t_q+t_{\rm wait}+\Delta\right].
\label{eq:readout-window}
\end{equation}

As a fixed baseline for the local distinguishability present before Hamiltonian switching, we evaluate the same observables over a window of width $\Delta$ immediately preceding the beginning of the quench-time scan,
\begin{equation}
t\in [100-\Delta,100].
\label{eq:sfim-baseline-window}
\end{equation}
This SFIM baseline is compared with the post-quench TFIM readout window,
\begin{equation}
t\in [t_q+t_{\rm wait},\,t_q+t_{\rm wait}+\Delta].
\label{eq:tfim-readout-window}
\end{equation}
Comparing these two windows allows us to quantify whether the Hamiltonian switch enhances the distinguishability of selected observables beyond the residual distinguishability already present during the preceding slanted-field evolution.

\subsection{Experimental realization on Rydberg atom arrays}
\label{subsec:rydberg}

The effective spin model in Eq.~\eqref{eq:ising-general} can be implemented in neutral-atom Rydberg arrays. We identify the two-level states with an atomic ground state \(|g\rangle\) and a Rydberg state \(|r\rangle\), and drive the transition with a near-resonant laser of Rabi frequency \(\Omega\) and detuning \(\delta\). In the rotating frame and within the rotating-wave approximation, the driven Rydberg Hamiltonian can be written as
\begin{equation}
\begin{aligned}
H_{\rm Ryd} = &
\sum_i \frac{\Omega}{2}\sigma_i^x
-\sum_i \frac{\delta}{2}\sigma_i^z\\
&+\sum_{i<j}V_{ij}
\left(\frac{1+\sigma_i^z}{2}\right)
\left(\frac{1+\sigma_j^z}{2}\right).
\label{eq:rydberg-hamiltonian}
\end{aligned}
\end{equation}
Here the Rydberg occupation is \(n_i=(1+\sigma_i^z)/2\).

For Rydberg states with a large Förster defect, the interaction is well described by a perturbative van der Waals form,
\begin{equation}
V_{ij}=\frac{C_6}{R_{ij}^6},
\end{equation}
where $R_{ij}$ is the distance between atoms $i$ and $j$. In a uniformly spaced linear chain, the next-nearest-neighbor interaction is smaller than the nearest-neighbor interaction by a factor $2^{-6}=1/64$. 

For a ring implementation, an $nS$ Rydberg state may be chosen so that the van der Waals interaction is approximately isotropic. The ratios between nearest- and longer-range interactions then depend on the chord distances of the finite ring and therefore on the total number of atoms. As $N$ increases, the local geometry approaches that of a linear chain, while the $R^{-6}$ dependence continues to suppress couplings beyond nearest neighbors. We therefore employ a nearest-neighbor Ising model as the idealized numerical description, while recognizing that a specific experimental implementation may include residual longer-range terms.

For the ring geometry, the nearest-neighbor interaction term becomes
\begin{align}
H_{int}&=V\sum_{i=1}^{N}
\frac{1+\sigma_i^z}{2}
\frac{1+\sigma_{i+1}^z}{2}\\
&=
\frac{V}{4}
\sum_{i=1}^{N}
\left(
1+\sigma_i^z+\sigma_{i+1}^z+\sigma_i^z\sigma_{i+1}^z
\right) \\
&=
\frac{V}{2}\sum_{i=1}^{N}\sigma_i^z
+
\frac{V}{4}\sum_{i=1}^{N}\sigma_i^z\sigma_{i+1}^z
+
\frac{NV}{4}\mathbb{I},
\label{eq:ring-rydberg-map}
\end{align}
with \(\sigma_{N+1}^\alpha\equiv\sigma_1^\alpha\). The term proportional to the identity only contributes a global phase and does not affect the dynamics.

Thus, for the ring geometry within the nearest-neighbor approximation, the driven Rydberg Hamiltonian maps onto the Ising form used in this work, with
\begin{equation}
B_x = \frac{\Omega}{2},
\qquad
B_z = \frac{V-\delta}{2},
\qquad
J = \frac{V}{4},
\label{eq:rydberg-ising-parameters}
\end{equation}
up to the spin convention used to define \(\sigma^z\). The effective longitudinal field is therefore controlled by both the laser detuning and the interaction-induced shift. The Hamiltonian-switching protocol considered here can be implemented by dynamically changing global laser parameters \(\Omega(t)\) and \(\delta(t)\), while leaving the atom geometry fixed.

Programmable Rydberg atom arrays also support site-resolved measurements of magnetization and spin--spin correlation functions. By applying coherent spin rotations before state-selective detection, these observables can be accessed in different measurement bases \cite{Ber17,Eba20,Scholl2021}.

\subsection{Numerical methods}

The dynamics were computed by exact state-vector evolution in the full $2^N$-dimensional Hilbert space using the unitary propagator $e^{-iHt}$. All calculations were performed without Hilbert-space truncation.

Candidate quench times were scanned over $Jt_q\in[100,110]$ with spacing $J\delta t_q=0.1$. For each candidate quench time, the observable trajectories entering the time-averaged signals and fluctuation widths were sampled with temporal spacing $J\delta t=1$. The optimized quench time $t_{q,O}^{*}$ was selected separately for each observable and system size by maximizing $\left|R_O^{\rm TFIM}(t_q)\right|$ over this scan. All reported values of $R_O^{\rm TFIM}$, $G_O^{\rm opt}$, $\mathcal{V}_O^{\rm opt}$, and $\mathcal{D}_O^{\rm opt}$ were evaluated using this common temporal sampling.

The representative $M_x^A(t)$ trajectories shown in Fig.~\ref{fig:mx-readout-trace} were evaluated with the finer temporal spacing $J\delta t=0.1$ to resolve their detailed time dependence. The half-chain von Neumann entropy and Hilbert--Schmidt distance under SFIM evolution were likewise evaluated with $J\delta t=0.1$, since these calculations require a single uninterrupted SFIM trajectory rather than a scan over quench times. Time-averaged quantities and temporal fluctuation widths were computed as discrete averages over the sampled points within the corresponding analysis windows. We verified for representative system sizes and observables that reducing the temporal sampling interval does not alter the qualitative channel hierarchy or finite-size conclusions.

\section{Subsystem observables and readout metrics}\label{sec:obs}

\subsection{Subsystem partition and reduced-state diagnostics}
\label{subsec:subsystem}
We focus on a contiguous subsystem \(A\) of \(N_A\) spins. 
Unless otherwise stated, \(A\) is chosen as a half-chain subsystem,
\[
A=\{1,2,\ldots,N_A\},
\qquad
N_A=\left\lfloor \frac{N}{2}\right\rfloor .
\]
For even \(N\), this is the usual half-chain partition, while for odd \(N\) the accessible subsystem contains $\frac{N-1}{2}$ spin. Observables are averaged over sites and bonds contained entirely within \(A\). This choice reflects the system--bath viewpoint introduced above: \(A\) is the accessible subsystem, while the complement \(\bar A\) acts as an internal bath.

For each evolved state \(|\psi_\pm(t)\rangle\) and its density matrix $
\rho^{(\pm)}(t) = |\psi_\pm(t)\rangle\langle\psi_\pm(t)|$, we define the reduced density matrix of subsystem \(A\) as
\begin{equation}
\rho_A^{(\pm)}(t)
=
\mathrm{Tr}_{\bar A} \left[ \rho^{(\pm)}(t) \right].
\label{eq:rhoA}
\end{equation}
We quantify entanglement between \(A\) and \(\bar A\) using the von Neumann entropy
\begin{equation}
S_A^{(\pm)}(t)
=
-\mathrm{Tr}
\left[
\rho_A^{(\pm)}(t)\log\rho_A^{(\pm)}(t)
\right].
\label{eq:von-neumann-entropy}
\end{equation}
Since the full state remains pure, \(S_A^{(\pm)}\) measures entanglement generated between the accessible subsystem and its internal bath.

\begin{figure}[t]
    \centering
    \includegraphics[width=\linewidth]{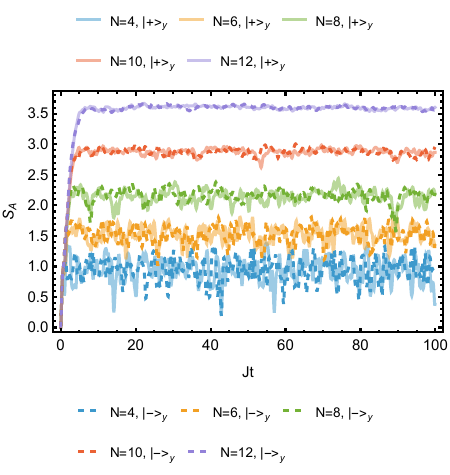}
    \includegraphics[width=\linewidth]{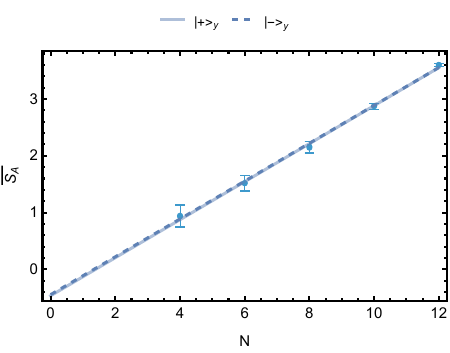}
    \caption{Half-chain entanglement growth under SFIM evolution. \textit{(top)} Von Neumann entropy $S_A^{(\pm)}(t)$ versus dimensionless time $Jt$ for several system sizes, starting from $|\pm y\rangle^{\otimes N}$. Solid transparent curves correspond to $|+y\rangle^{\otimes N}$, while dashed opaque curves correspond to $|-y\rangle^{\otimes N}$. The entropy grows rapidly from zero and subsequently fluctuates around a finite late-time value. \textit{(bottom)} Late-time averaged entropy $\overline{S_A^{(\pm)}}$, computed over $30\le Jt\le100$, as a function of subsystem size $N_A$. The data for the two initial states nearly coincide. Solid transparent and dashed opaque lines show separate linear fits for the $|+y\rangle^{\otimes N}$ and $|-y\rangle^{\otimes N}$ data, respectively.}
    \label{fig:half-chain-entropy}
\end{figure}

Figure~\ref{fig:half-chain-entropy} \textit{(top)} shows the half-chain entropy during evolution under the slanted-field Ising Hamiltonian. Starting from the product states \(|\pm y\rangle^{\otimes N}\), \(S_A^{(\pm)}(t)\) grows rapidly from zero and then fluctuates around a finite late-time value. The two initial states exhibit nearly identical entropy growth on the scale shown, indicating that the SFIM stage generates entanglement between the half-chain subsystem $A$ and its complement $\bar A$ at comparable rates for both initial states. The late-time averaged entropy increases approximately linearly with subsystem size over the finite sizes studied, as shown in Fig.~\ref{fig:half-chain-entropy} \textit{(bottom)}, consistent with volume-law-like entanglement growth.

\begin{figure}[h]
    \centering
    \includegraphics[width=\linewidth]{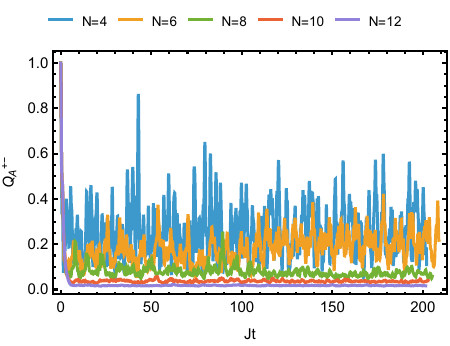}
    \includegraphics[width=\linewidth]{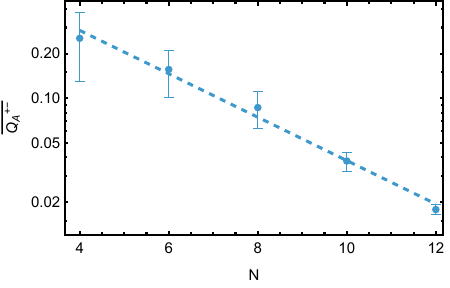}
    \caption{ Half-chain reduced-state separation under SFIM evolution. \textit{(top)} Half squared Hilbert--Schmidt distance $D_{HS,A}^{+-}(t)$ between the half-chain reduced density matrices evolved from $|+y\rangle^{\otimes N}$ and $|-y\rangle^{\otimes N}$. The distance rapidly decreases, indicating that the separation between the two reduced states becomes strongly suppressed within subsystem $A$ according to this quadratic diagnostic. \textit{(bottom)} Late-time averaged residual distance $\overline{D_{HS,A}^{+-}}$, computed over $30\le Jt\le100$, as a function of system size $N$. The residual distance decreases rapidly with $N$; the dashed line is a linear fit to $\log\left(\overline{D_{HS,A}^{+-}}\right)$ over the finite-size range $N=4,6,8,10,12$. }
    \label{fig:half-chain-hs}
\end{figure}

To quantify the separation between the two initial states within subsystem $A$, we use the half squared Hilbert--Schmidt distance, defined as
\begin{equation}
D_{HS,A}^{+-}(t) = \frac{1}{2} \mathrm{Tr} \left[\left(\rho_A^{(+)}(t)-\rho_A^{(-)}(t)\right)^2\right].
\label{eq:hs-distance}
\end{equation}
This quantity is a basis-independent quadratic measure of reduced-state separation and provides a computationally efficient diagnostic of whether the two reduced density matrices become locally similar under the slanted-field evolution.

Figure~\ref{fig:half-chain-hs} shows that the SFIM evolution strongly suppresses the separation between the two reduced states. For the initial orthogonal product states used here, the reduced states on $A$ are pure and orthogonal at $t=0$, so the normalization in Eq.~\eqref{eq:hs-distance} gives $D_{HS,A}^{+-}(0)=1$. Under SFIM evolution, $D_{HS,A}^{+-}(t)$ rapidly decreases from this initial value and settles to a finite-size-dependent residual value. This residual value decreases with increasing $N$, indicating that the reduced density matrices generated from $|+y\rangle^{\otimes N}$ and $|-y\rangle^{\otimes N}$ become increasingly close according to the Hilbert--Schmidt diagnostic within the half-chain subsystem.

On a semilog scale, the residual half-chain separation is approximately linear over $N=4$ to $12$ consistent with rapid, roughly exponential suppression over the accessible finite-size range. We do not infer an asymptotic scaling law from these sizes. The nonzero residual value of $D_{HS,A}^{+-}$ shows that the two reduced states are not exactly identical at finite size. Thus, the SFIM stage does not erase the initial-state dependence even within the subsystem. Rather, it strongly suppresses its local accessibility. Because the corresponding global states remain orthogonal under unitary evolution, the global initial-state distinction remains preserved, including in degrees of freedom outside $A$ and in correlations across the bipartition.

These diagnostics establish the SFIM stage as a delocalizing evolution that strongly suppresses the separation of the reduced subsystem states. We next ask whether a transverse-field readout stage can enhance the sensitivity of selected one- and two-body observables to the initial state.

\subsection{Observable readout after transverse-field switching}
\label{subsec:observable-readout}

We now quantify the Hamiltonian-switching response at the level of experimentally accessible local and two-body observables. For each initial state $|\psi_\pm(0)\rangle$, we define
\begin{equation}
O_\pm(t)
=
\langle \psi_\pm(t)|O|\psi_\pm(t)\rangle ,
\end{equation}
where $O$ denotes the observable of interest. 

We focus on the following local observables within subsystem $A$: the subsystem-averaged magnetization and the mean nearest-neighbor correlations. The subsystem-averaged magnetization is
\begin{equation}
M_\alpha^A(t)
=
\frac{1}{N_A}
\sum_{i\in A}
\langle \sigma_i^\alpha(t)\rangle ,
\qquad \alpha=x,z,
\end{equation}
and the nearest-neighbor correlation averaged averaged over bonds contained within $A$ is
\begin{equation}
C_{\alpha\alpha}^A(t)
=
\frac{1}{N_A -1}
\sum_{i=1}^{N_A-1}
\langle \sigma_i^\alpha(t)\sigma_{i+1}^\alpha(t)\rangle ,
\qquad \alpha=x,z.
\end{equation}
Here the bond $(i,i+1)$ follows the ring ordering.

\begin{figure}[h]
    \centering
    \includegraphics[width=\linewidth]
    {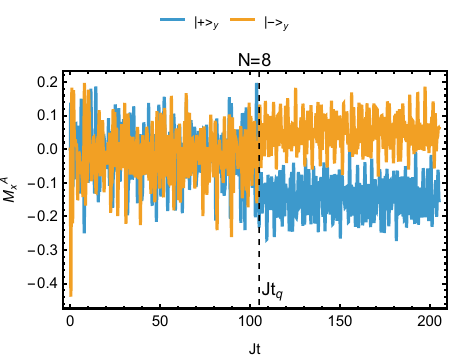} 
    \includegraphics[width=\linewidth]
    {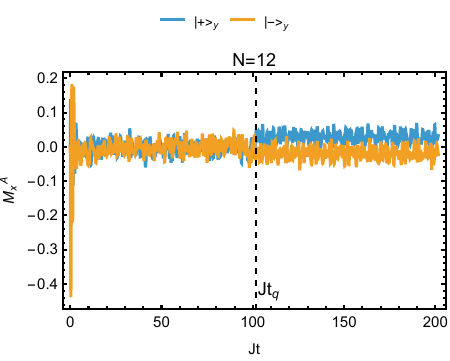} 
    \caption{
    Representative observable readout after Hamiltonian switching for
    $N=8$ and $N=12$.
    The subsystem-averaged transverse magnetization $M_x^A(t)$ is shown
    for the two initial states $|+y\rangle^{\otimes N}$ in blue and
    $|-y\rangle^{\otimes N}$ in yellow.
    \textit{(top)} For $N=8$, the two signals have nearly identical
    means and fluctuation amplitudes during the SFIM evolution, and
    become nearly nonoverlapping after the switch to TFIM.
    \textit{(bottom)} For $N=12$, the overall fluctuation amplitude is
    smaller, but the two post-quench trajectories remain visibly
    separated.
    The vertical dashed lines indicate the respective quench times,
    after which the system evolves under the transverse-field Ising
    Hamiltonian.}
    \label{fig:mx-readout-trace}
\end{figure}

To illustrate the readout effect at the level of a directly measurable
observable, Fig.~\ref{fig:mx-readout-trace} shows representative time
traces of the subsystem-averaged transverse magnetization $M_x^A(t)$
for $N=8$ and $N=12$. During the late-time SFIM stage, the two
trajectories are strongly intermixed. Importantly, they not only
oscillate around nearly identical mean values, but also exhibit similar
fluctuation amplitudes for the two initial states.

For both system sizes, the two SFIM trajectories have nearly identical means and comparable fluctuation widths, making the initial states difficult to distinguish through $M_x^A$ before the switch. 

After the TFIM switch, the trajectories separate into clearly distinct dynamical bands. For $N=8$, the two post-quench traces are nearly nonoverlapping over the readout window, indicating strong initial-state sensitivity in the $M_x^A$ channel. 

Although both the absolute separation and the fluctuation amplitudes decrease with increasing system size, the post-quench bands remain visibly separated at $N=12$. These representative traces therefore show that the reduction of the raw observable scale does not by itself imply a corresponding loss of dynamical distinguishability.

These traces provide a concrete illustration of the central mechanism of the protocol. The Hamiltonian switch does not reverse the preceding many-body evolution. Instead, local observables that had become only weakly sensitive to the distinction between the two initial states during the SFIM stage regain strong initial-state sensitivity under the TFIM dynamics. In selected measurement channels, the observable signals associated with the two orthogonal initial states can form nearly nonoverlapping dynamical bands, allowing the states to be distinguished directly from their time traces. This separation is particularly pronounced for $M_x^A$ at $N=8$ and remains clearly observable at $N=12$, despite the reduction of the absolute observable amplitude with increasing system size.

The TFIM stage is used as an observable-selective readout: it enhances the sensitivity of selected one- and two-body measurement channels to the initial state. We do not interpret this enhancement as a general recovery of the half-chain reduced state. Representative calculations of the half-chain Hilbert--Schmidt distance show either an increase or a decrease after the switch, depending on the system size and quench condition. These limited calculations therefore do not establish a systematic trend in the post-quench reduced-state distinguishability.

Qualitatively and quantitatively similar behavior is observed in the other measurement channels considered here. Under SFIM evolution, the signals generated from the two initial states are strongly intermixed,
with similar mean values and comparable temporal fluctuation widths.
After the switch to TFIM, enhanced separation of the time-averaged
signals is also observed for $C_{zz}^A$, $C_{xx}^A$, and $M_z^A$. The enhancement is observable-dependent and is generally weaker for $M_z^A$ than for $M_x^A$ and the two correlation channels.

\subsection{Quench-time dependence and finite-size readout visibility}
\label{subsec:gain-scaling}

We now quantify the Hamiltonian-switching response by comparing the
observable signals generated from the two initial states before and
after the switch. After a quench at time $t_q$, the system evolves
under the transverse-field Ising Hamiltonian for a waiting time
$t_{\rm wait}$. The observable is then averaged over a readout window
of width $\Delta$:
\begin{equation}
\mu_{O,\pm}^{\rm TFIM}(t_q)
=
\frac{1}{\Delta}
\int_{t_q+t_{\rm wait}}^{t_q+t_{\rm wait}+\Delta}
O_\pm(t)\,dt .
\label{eq:tfim-readout-average}
\end{equation}

We also define, for comparison, a fixed SFIM reference by averaging the same observable over the final window of the initial SFIM evolution, before any Hamiltonian switch is applied. Because the quench-time scan begins at $Jt_q=100$,
we define
\begin{equation}
\mu_{O,\pm}^{\rm SFIM}
=
\frac{1}{\Delta}
\int_{100-\Delta}^{100}
O_\pm(t)\,dt .
\label{eq:sfim-baseline-average}
\end{equation}
This reference characterizes the residual separation already visible
in the selected measurement channel during the late-time SFIM
evolution.

The signed post-quench and pre-quench responses are
\begin{equation}
R_O^{\rm TFIM}(t_q)
=
\mu_{O,+}^{\rm TFIM}(t_q)
-
\mu_{O,-}^{\rm TFIM}(t_q),
\label{eq:tfim-response}
\end{equation}
and
\begin{equation}
R_O^{\rm SFIM}
=
\mu_{O,+}^{\rm SFIM}
-
\mu_{O,-}^{\rm SFIM}.
\label{eq:sfim-response}
\end{equation}
The sign specifies the orientation of the separation between the signals generated from the two initial states, while $|R_O^{\rm TFIM}(t_q)|$ and $|R_O^{\rm SFIM}|$ give the corresponding separation magnitudes.

\begin{figure}[h]
    \centering
    \includegraphics[width=\linewidth]
    {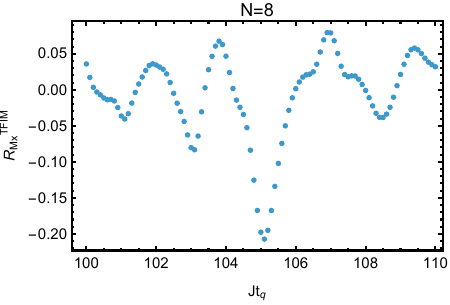}
    \caption{
    Quench-time dependence of the signed post-quench response
    $R_{M_x^A}^{\rm TFIM}(t_q)$ for $N=8$, $t_{\rm wait}=30$ and $\Delta=70$.}
    \label{fig:response-vs-tq}
\end{figure}

Figure~\ref{fig:response-vs-tq} shows the quench-time dependence of the signed separation $R_{M_x^A}^{\rm TFIM}(t_q)$ between the two time-averaged $M_x^A$ signals for $N=8$. The response changes
both magnitude and sign as $t_q$ is varied. At multiple quench times across this window, values of $\left|R_{M_x^A}^{\rm TFIM}(t_q)\right|$ substantially larger than the fixed SFIM baseline, $\left|R_{M_x^A}^{\rm SFIM}\right|=3.73\times10^{-4}$, are obtained. This demonstrates that the
TFIM readout depends on the many-body state reached during the preceding SFIM evolution, rather than only on the choice of the final Hamiltonian. Similar quench-time-dependent structure is observed for
$M_z^A$, $C_{xx}^A$, and $C_{zz}^A$, although the amplitudes and
detailed oscillation patterns are observable-dependent.

To test whether the enhanced readout separation found near $Jt_q\in[100,110]$ is washed out at substantially longer time scales by slow thermalization or other long-time relaxation processes, we also evaluate the response in the separated windows
$Jt_q\in[1000,1010]$, $Jt_q\in[10000,10010]$, and
$Jt_q\in[100000,100010]$ for the representative system size $N=8$.
This is not a continuous scan over all possible quench times. Rather,
it tests whether responses larger than the fixed SFIM separation recur
after substantially longer SFIM evolution.

\begin{figure}[h]
    \centering
    \includegraphics[width=\linewidth]
    {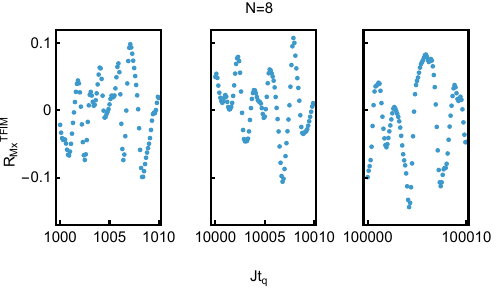}
    \caption{
    Long-time check of the signed post-quench response
    $R_{M_x^A}^{\rm TFIM}(t_q)$ for $N=8$ over several separated
    quench-time windows of width $10$.
    The fixed SFIM response is shown as a horizontal reference.
    Post-quench responses with magnitude largely exceeding the SFIM separation
    recur in the later windows, although the scan is fragmented rather
    than continuous in $t_q$.}
    \label{fig:fragmented-response-windows}
\end{figure}

Figure~\ref{fig:fragmented-response-windows} shows that the response
remains strongly quench-time dependent within each widely separated
interval. Values satisfying $|R_{M_x^A}^{\rm TFIM}(t_q)|>>|R_{M_x^A}^{\rm SFIM}|$
recur after orders-of-magnitude longer SFIM evolution times. Moreover, no clear systematic reduction in the attainable magnitude of the separation $\left|R_{M_x^A}^{\rm TFIM}(t_q)\right|$ is evident across the sampled long-time windows. The enhancement observed near $Jt_q\in[100,110]$ is therefore not confined to the first sampled late-time window, but recurs as part of the long-time coherent dynamics.

Because the response varies strongly with the switching time, we
identify, for each observable and system size, the quench time that
maximizes the absolute post-quench separation within the finite scan:
\begin{equation}
t_{q,O}^{*}
=
\underset{t_q\in[100,110]}{\operatorname{arg\,max}}
\,
\left|R_O^{\rm TFIM}(t_q)\right|.
\label{eq:optimized-quench-time}
\end{equation}
The corresponding optimized absolute separation is
\begin{equation}
\begin{aligned}
\left|R_O^{\rm TFIM}\right|_{\rm opt}
&=
\left|
R_O^{\rm TFIM}\left(t_{q,O}^{*}\right)
\right|\\
&=
\max_{t_q\in[100,110]}
\left|R_O^{\rm TFIM}(t_q)\right|.
\label{eq:optimized-raw-separation}
\end{aligned}
\end{equation}
This is a finite-window optimum over $Jt_q\in[100,110]$, rather than a
global optimum over all possible switching times. A broader scan could
reveal additional peaks or sign changes.

To determine how strongly the Hamiltonian switch enhances the mean
separation relative to that already present during the SFIM dynamics,
we define the optimized readout gain
\begin{equation}
G_O^{\rm opt}
=
\frac{
\left|R_O^{\rm TFIM}\right|_{\rm opt}
}{
\left|R_O^{\rm SFIM}\right|
}.
\label{eq:optimized-gain}
\end{equation}
Values satisfying $G_O^{\rm opt}>1$ indicate that the strongest
post-quench separation found within the scan exceeds the mean
separation already present during the fixed SFIM reference window.
Thus, even when the instantaneous trajectories partially overlap
because of temporal fluctuations, the two initial states can remain
distinguishable through their different time-averaged observable
signals. The Hamiltonian switch therefore enhances the information
accessible in the mean readout relative to the preceding SFIM
dynamics.

To compare this mean separation with the characteristic temporal
fluctuations present before the switch, we define the SFIM fluctuation
widths
\begin{equation}
\sigma_{O,\pm}^{\rm SFIM}
=
\left[
\frac{1}{\Delta}
\int_{100-\Delta}^{100}
\left(
O_\pm(t)-\mu_{O,\pm}^{\rm SFIM}
\right)^2
dt
\right]^{1/2}.
\label{eq:sfim-standard-deviation}
\end{equation}
Although the two SFIM widths are nearly identical in the present
calculations, we combine them symmetrically through the pooled
fluctuation scale
\begin{equation}
\sigma_{O,\rm pool}^{\rm SFIM}
=
\sqrt{
\frac{
\left(\sigma_{O,+}^{\rm SFIM}\right)^2
+
\left(\sigma_{O,-}^{\rm SFIM}\right)^2
}{2}
}.
\label{eq:sfim-pooled-standard-deviation}
\end{equation}

A large separation relative to the SFIM fluctuation scale does not by
itself guarantee that the two post-quench trajectories are well
separated, because the TFIM dynamics can produce a different
fluctuation amplitude. We therefore also evaluate the TFIM fluctuation
widths at the same optimized quench time $t_{q,O}^{*}$ for each observable $O$:
\begin{equation}
\begin{aligned}
&\sigma_{O,\pm}^{\rm TFIM,*}
=\\
&\left[
\frac{1}{\Delta}
\int_{t_{q,O}^{*}+t_{\rm wait}}
^{t_{q,O}^{*}+t_{\rm wait}+\Delta}
\left[
O_\pm(t)
-
\mu_{O,\pm}^{\rm TFIM}
\left(t_{q,O}^{*}\right)
\right]^2
dt
\right]^{1/2}.
\end{aligned}
\label{eq:optimized-tfim-standard-deviation}
\end{equation}

\begin{figure}
    \centering
    \includegraphics[width=\linewidth]
    {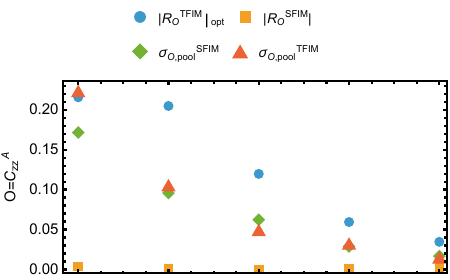}
    \includegraphics[width=\linewidth]
    {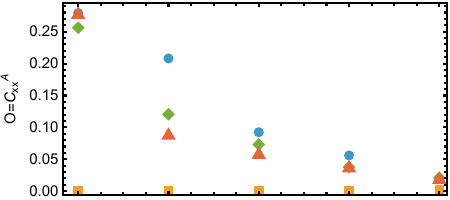}
    \includegraphics[width=\linewidth]
    {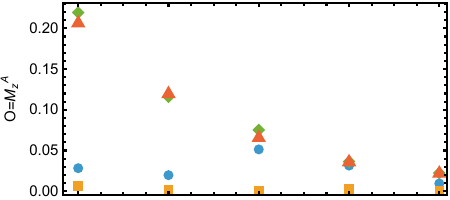}
    \includegraphics[width=\linewidth]
    {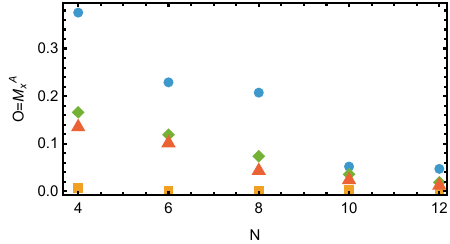}
    \caption{
    Finite-size dependence of the absolute readout and dynamical scales
    for the even system sizes $N=4,6,8,10,12$, using $\Delta=70$ and
    $Jt_q\in[100,110]$.
    The four panels correspond to
    $C_{zz}^A$,  $C_{xx}^A$, $M_z^A$, and  $M_x^A$.
    Distinct marker shapes show the optimized TFIM mean separation
    $\left|R_O^{\rm TFIM}\right|_{\rm opt}$, the fixed SFIM mean
    separation $\left|R_O^{\rm SFIM}\right|$, the pooled SFIM
    fluctuation width $\sigma_{O,\rm pool}^{\rm SFIM}$, and the pooled
    TFIM fluctuation width $\sigma_{O,\rm pool}^{\rm TFIM,*}$ evaluated
    at the same optimized quench time.
    Odd system sizes are reported separately in the Appendix.}
    \label{fig:absolute-readout-scales}
\end{figure}
The two post-quench fluctuation widths are not required to be equal.
We combine them through the pooled TFIM fluctuation scale
\begin{equation}
\sigma_{O,\rm pool}^{\rm TFIM,*}
=
\sqrt{
\frac{
\left(\sigma_{O,+}^{\rm TFIM,*}\right)^2
+
\left(\sigma_{O,-}^{\rm TFIM,*}\right)^2
}{2}
}.
\label{eq:optimized-tfim-pooled-standard-deviation}
\end{equation}
Figure~\ref{fig:absolute-readout-scales} shows that the optimized
post-quench response generally decreases in absolute magnitude as the
system size increases. However, this decrease occurs together with a
substantial reduction of both the SFIM and TFIM fluctuation widths.
The shrinking raw response therefore reflects, at least in part, the
decreasing dynamical amplitude of the subsystem-averaged observables,
rather than by itself demonstrating a proportional loss of readout
distinguishability.

The fixed SFIM mean separation
$\left|R_O^{\rm SFIM}\right|$ is substantially smaller than the
corresponding SFIM fluctuation width in most cases, quantitatively
confirming that the two initial-state signals are strongly intermixed
before the Hamiltonian switch. In contrast,
$\left|R_O^{\rm TFIM}\right|_{\rm opt}$ is typically comparable to or
larger than the relevant dynamical fluctuation scales in the stronger
measurement channels.

Consistent with this observation, $G_O^{\rm opt}>1$ for every
measurement channel and system size investigated. The Hamiltonian
switch therefore produces a larger mean separation than the residual
SFIM baseline throughout the accessible range. However,
$G_O^{\rm opt}$ is strongly nonmonotonic because its denominator can
become extremely small. We consequently retain the gain as a measure
of enhancement relative to the residual SFIM mean separation, but do
not use it as the primary finite-size scaling diagnostic.

We next define two standardized mean-separation measures. The first is
the optimized SFIM-normalized readout visibility
\begin{equation}
\mathcal{V}_O^{\rm opt}
=
\frac{
\left|R_O^{\rm TFIM}\right|_{\rm opt}
}{
\sigma_{O,\rm pool}^{\rm SFIM}
}.
\label{eq:sfim-normalized-tfim-separation}
\end{equation}
This quantity has the structure of a Glass's-$\Delta$-type standardized
mean difference: the optimized post-quench mean separation is
normalized by the fluctuation scale of the reference SFIM dynamics.
Here, the two SFIM widths are pooled symmetrically because neither
initial state serves as a unique control trajectory.

The second is the optimized TFIM-normalized dynamical-band separation
\begin{equation}
\mathcal{D}_O^{\rm opt}
=
\frac{
\left|R_O^{\rm TFIM}\right|_{\rm opt}
}{
\sigma_{O,\rm pool}^{\rm TFIM,*}
}.
\label{eq:optimized-tfim-band-separation}
\end{equation}
This quantity has the structure of a Cohen's-$d$-type standardized mean
difference: the difference between the two TFIM mean signals is
normalized by their pooled post-quench fluctuation width.

The correspondence with Glass's $\Delta$ and Cohen's $d$ is mathematical rather than inferential. The widths entering Eqs.~\eqref{eq:sfim-normalized-tfim-separation} and \eqref{eq:optimized-tfim-band-separation} describe coherent temporal fluctuations of deterministic many-body trajectories, rather than independent statistical samples. These quantities therefore provide dimensionless measures of mean-signal separation relative to the relevant dynamical fluctuation scales.

Importantly, $\mathcal{D}_O^{\rm opt}$ is evaluated at the same
$t_{q,O}^{*}$ that maximizes
$\left|R_O^{\rm TFIM}(t_q)\right|$. We do not optimize the denominator
or $\mathcal{D}_O^{\rm opt}$ independently. Thus, both standardized
measures characterize the same selected readout condition.

\begin{figure}[h]
    \centering
    \includegraphics[width=\linewidth]
    {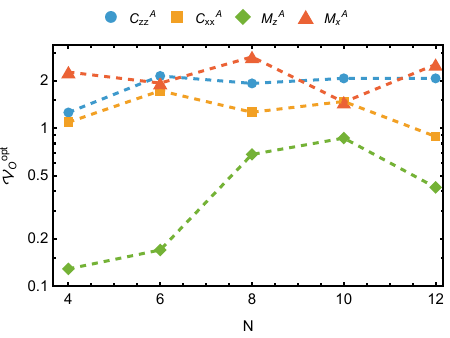}
    \includegraphics[width=\linewidth]
    {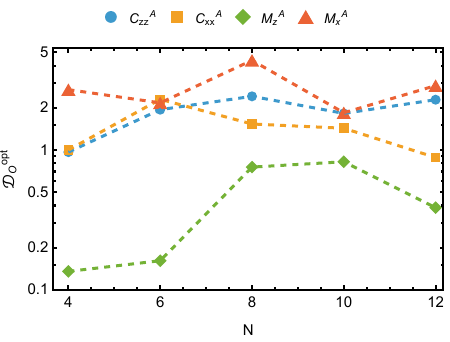}
    \caption{Finite-size dependence of the two standardized readout measures for the even system sizes $N=4,6,8,10,12$. \textit{(top)} SFIM-normalized readout visibility $\mathcal{V}_O^{\rm opt}$, a Glass's-$\Delta$-type measure using the pooled SFIM fluctuation width as the reference scale. \textit{(bottom)} TFIM-normalized dynamical-band separation $\mathcal{D}_O^{\rm opt}$, a Cohen's-$d$-type measure using the pooled TFIM fluctuation width at the optimized quench time. The observables are $M_x^A$, $M_z^A$, $C_{xx}^A$, and $C_{zz}^A$. Both panels use a logarithmic vertical scale. Dashed lines connecting successive system sizes are included as guides to the eye to visualize the finite-size trends and the absence of a systematic large-size suppression in the strongest channels. Odd system sizes are reported separately in the Appendix.}
    \label{fig:standardized-readout-scaling}
\end{figure}

Because $\mathcal{V}_O^{\rm opt}$ and
$\mathcal{D}_O^{\rm opt}$ are standardized mean-separation measures, their values should be interpreted continuously rather than through a strict threshold at unity. Values below one can still represent a
substantial displacement between the two dynamical bands, while values well above one indicate increasingly strong separation relative to the
relevant fluctuation scale.

Figure~\ref{fig:standardized-readout-scaling}\textit{(top)} shows that the SFIM-normalized visibility is robust in the strongest measurement channels. For $M_x^A$, $\mathcal{V}_O^{\rm opt}$ lies between approximately $1.46$ and $2.83$ across the accessible even sizes, while for $C_{zz}^A$ it remains between approximately $1.26$ and $2.14$. Thus, in both channels, the optimized post-quench mean separation is consistently larger than the characteristic fluctuation amplitude of the preceding SFIM dynamics, with no systematic decrease over the accessible size range. The $C_{xx}^A$ channel also retains a substantial standardized separation, ranging from approximately $0.89$ to $1.73$, and remains close to one even at $N=12$. The $M_z^A$ channel is weaker, but still exhibits an observable-dependent enhancement, reaching approximately $0.87$ at $N=10$.

The TFIM-normalized dynamical-band separation in Fig.~\ref{fig:standardized-readout-scaling}\textit{(bottom)} provides a more direct assessment of the distinguishability of the post-quench time traces themselves. The $M_x^A$ channel exhibits particularly strong separation: $\mathcal{D}_{M_x^A}^{\rm opt}$ ranges from approximately $1.84$ to $4.37$ across the even sizes. At $N=8$, the two means are separated by more than four pooled TFIM fluctuation widths, consistent with the nearly nonoverlapping dynamical bands in Fig.~\ref{fig:mx-readout-trace}. Even at $N=12$, the separation remains approximately $2.88$ pooled widths. The $C_{zz}^A$ channel is similarly robust. Apart from the smallest system, its post-quench means remain separated by approximately $1.82$--$2.42$ pooled TFIM widths, reaching $2.28$ at $N=12$. The $C_{xx}^A$ channel exhibits moderate to strong separation for most sizes, with values as large as $2.30$ and a value of approximately $0.88$ at $N=12$. Although this final value is below unity, it still corresponds to a mean displacement comparable to the characteristic width of the post-quench trajectories and should not be interpreted as an absence of readout. The $M_z^A$ channel remains the least strongly resolved, with $\mathcal{D}_{M_z^A}^{\rm opt}$ below one throughout the accessible even-size range.

Although the raw separation and fluctuation amplitudes decrease with system size, the standardized readout shows no corresponding systematic suppression in the strongest channels. The $M_x^A$ and $C_{zz}^A$ signals remain separated by multiple intrinsic fluctuation widths through $N=12$, while $C_{xx}^A$ remains of order unity. The shrinking absolute response therefore reflects the simultaneous narrowing of the pre- and post-quench dynamical bands rather than a simple finite-size collapse of the readout.

The quantities introduced above have complementary roles:
\begin{enumerate}
    \item $\left|R_O^{\rm TFIM}\right|_{\rm opt}$ and $\left|R_O^{\rm SFIM}\right|$ give, respectively, the strongest post-quench mean separation found within the scanned quench-time window and the residual mean separation already present during the fixed SFIM reference window.
    
    \item $G_O^{\rm opt}$ measures the enhancement of the optimized
    post-quench separation relative to this residual SFIM mean
    separation.

    \item $\mathcal{V}_O^{\rm opt}$ is a
    Glass's-$\Delta$-type measure of the post-quench separation relative
    to the characteristic fluctuation scale of the preceding SFIM
    dynamics.

    \item $\mathcal{D}_O^{\rm opt}$ is a Cohen's-$d$-type measure of the
    post-quench separation relative to the pooled fluctuation width of
    the TFIM readout trajectories themselves.
\end{enumerate}

We do not infer thermodynamic behavior from these finite systems. The
optimization is restricted to the interval $Jt_q\in[100,110]$, and
additional quench times could reveal larger or smaller responses.
Across all observables considered, neither $\mathcal{V}_O^{\rm opt}$ nor $\mathcal{D}_O^{\rm opt}$ exhibits a clear monotonic decrease with increasing system size over the accessible range, although the magnitude of the standardized separation remains strongly observable dependent. Odd system sizes display the same qualitative absence of a
simple monotonic trend and are presented separately in the Appendix. The results therefore provide no evidence of a systematic finite-size collapse over $N=4$--$12$, while not establishing the asymptotic behavior at larger sizes.

Testing whether the standardized visibility and dynamical-band
separation persist at substantially larger system sizes requires
methods beyond desktop-scale exact state-vector simulation.
Programmable Rydberg atom arrays provide a natural platform for
extending this test and determining whether the readout separation
eventually decreases or remains finite beyond the accessible numerical
range.

\section{Conclusion and outlook}
\label{sec:conclusion}

We have introduced Hamiltonian switching as a readout method for
converting coherently stored many-body information into accessible
local signals. Two orthogonal product states first evolve under a
slanted-field Ising Hamiltonian, where entanglement grows rapidly and
their distinguishability within a half-chain subsystem becomes strongly
suppressed. A subsequent switch to the transverse-field Ising
Hamiltonian produces a markedly different dynamical response. Unlike
an echo or time-reversal protocol, this procedure neither reconstructs
the initial state nor attempts to invert the preceding evolution.
Instead, a simple global change of Hamiltonian maps part of the hidden
initial-state dependence into measurable one- and two-body observables.

During the SFIM stage, the half-chain entropy exhibits
volume-law-like finite-size scaling, the Hilbert--Schmidt distance
between the reduced states becomes small, and the observable
trajectories generated from the two initial states are strongly
intermixed. Their means and temporal fluctuation widths are nearly
identical, making the initial states difficult to distinguish in the
chosen local measurement channels. After the switch, the same
observables oscillate around distinct mean values. Across all tested observables and system sizes, the Hamiltonian switch enhances the time-averaged readout separation relative to the corresponding SFIM baseline, although the strength of the resulting readout remains strongly observable dependent. In the most sensitive channels, this enhancement also remains substantial when the post-quench fluctuations themselves are taken into account. For $M_x^A$ at $N=8$, the resulting dynamical bands are nearly nonoverlapping, while at $N=12$ their means remain separated by approximately three pooled TFIM fluctuation widths. The observed enhancement is therefore not merely a consequence of dividing by an anomalously small residual SFIM mean difference, but represents a genuine increase in the accessibility of the initial-state dependence through selected local and two-body observables.

The absolute response decreases with increasing system size, but so do
the characteristic SFIM and TFIM fluctuation amplitudes of the
subsystem-averaged observables. Consequently, the standardized readout
measures provide a substantially more informative finite-size
comparison than the raw response alone. The SFIM-normalized visibility,
which has the structure of a Glass's-$\Delta$-type standardized mean
difference, and the TFIM-normalized dynamical-band separation, which
has the structure of a Cohen's-$d$-type measure, remain clearly above
unity through $N=12$ for $M_x^A$ and $C_{zz}^A$. The $C_{xx}^A$
channel remains of order unity, whereas $M_z^A$ provides a weaker
readout. Thus, the strongest measurement channels retain a separation
comparable to or larger than their intrinsic dynamical widths across
the full range accessible to exact simulation.

The readout is strongly dependent on the quench time, including changes in both magnitude and sign. Enhanced readout also recurs in widely separated quench-time
windows extending to SFIM evolution times of order $10^5$. The enhancement is therefore not confined to the first sampled late-time window, but recurs at SFIM evolution times extending to order $10^5$, demonstrating that favorable readout conditions remain available within the long-time coherent dynamics.

This is the central physical significance of the phenomenon. Local equilibration under one Hamiltonian can make two states appear nearly indistinguishable in selected observables, creating the appearance that the information distinguishing them has been lost. In the closed many-body system considered here, however, that information remains encoded in the global state and in correlations extending beyond the accessible subsystem. A change in the global Hamiltonian parameters can redirect part of this hidden initial-state dependence into local measurement channels. Hamiltonian switching therefore acts as a many-body readout transformation: it changes which components of the stored information become visible in experimentally accessible observables. The emergence of large local signal separation without microscopic reversal shows that an apparent loss of information can instead reflect a loss of accessibility within the chosen measurement channels, and that hidden many-body memory can be actively redirected rather than merely passively detected.

Approximate numerical methods \cite{Vidal2004,Hae2011,Hae2016,PAE19,NAN25}, including tensor-network and related many-body approaches, may extend the accessible system sizes beyond exact state-vector simulation. However, their accuracy cannot generally be assumed in the long-time, entanglement-rich regime considered here, where controlled convergence may become difficult. 

Programmable Rydberg atom arrays therefore provide a more direct route to testing the mechanism at larger system sizes \cite{Eba20,Scholl2021,Ber17}. Product-state preparation, tunable transverse and longitudinal fields, rapid Hamiltonian switching, and site-resolved measurements of magnetization and correlations are all native capabilities of these platforms. Experiments can determine whether the standardized readout separation persists at larger atom numbers and test its robustness to long-range interactions, decoherence, preparation errors, and control noise. On the theoretical side, the single switch considered here can be generalized to optimized sequences of fields, detunings, and interaction strengths, turning Hamiltonian design into a practical tool for selecting how and where many-body information becomes observable.

\acknowledgements
The author thanks Alioscia Hamma for helpful discussions and comments on the finite-size analysis. This work was partially supported by U.S. National Science Foundation (NSF) Grant No. OSI-2328774. OpenAI ChatGPT and Undermind was used to assist in language revision, sentence formulation, literature and bibliography searches, and manuscript structuring. All simulations and calculations, all scientific ideas and conclusions were made by the author. The author verified all the cited sources, and reviewed all statements and resulting text.

\appendix

\section{Boundary-condition check}
\label{app:open-chain}

The main-text calculations use a ring geometry so that every site has the same number of nearest neighbors and the interaction-induced longitudinal field is spatially uniform. To check that the observed readout behavior is not a boundary artifact, we repeat representative calculations using an open chain, where the two edge spins have only one nearest neighbor.

\begin{figure}[h]
    \centering
    \includegraphics[width=\linewidth]{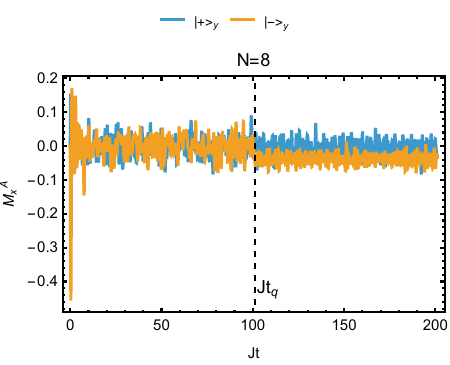}
    \includegraphics[width=\linewidth]{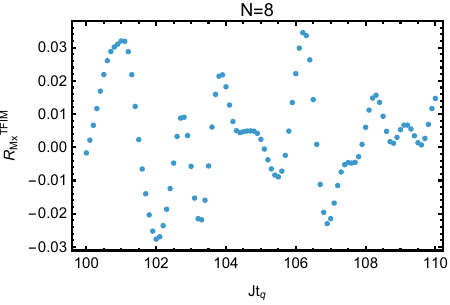}
    \caption{
    Boundary-condition check using an open chain for $N=8$. \textit{(top)} Representative time traces of $M_x^A(t)$ after Hamiltonian switching at $t_q=101$. \textit{(bottom)} Quench-time dependence of the signed mean separation $R_{M_x^A}^{\rm TFIM}(t_q)$ in the open-chain geometry. The post-quench dynamical-band separation and its oscillatory quench-time dependence persist for open boundaries, indicating that the readout effect is not specific to the periodic ring geometry used in the main text.
    }
    \label{fig:open-chain-check}
\end{figure}

Figure~\ref{fig:open-chain-check} presents an open-chain check for $N=8$ using the $M_x^A$ readout channel. The time traces show that the qualitative post-quench enhancement of the separation between the two initial states persists with open boundaries, while $R_{M_x^A}^{\rm TFIM}(t_q)$ retains a pronounced oscillatory dependence on the quench time. Because the boundary conditions alter the spectrum and detailed dynamics, we do not compare the quantitative readout metrics directly with those of the periodic ring. The open-chain result is instead used to establish that the qualitative Hamiltonian-switching readout mechanism is not specific to the ring geometry.

\section{Odd system sizes}
\label{app:odd-system-sizes}

The main text presents the finite-size analysis for even system sizes,
for which the periodic chain can be divided into two subsystems of equal
size. Here, we report the corresponding results for odd system sizes
$N=5,7,9,11$.

For odd $N$, the accessible subsystem is defined using the same
convention as in the main text,
\begin{equation}
N_A=\left\lfloor\frac{N}{2}\right\rfloor,
\qquad
A=\{1,\ldots,N_A\}.
\end{equation}
The complementary subsystem therefore contains one additional spin.
Because the subsystem dimensions and boundary-to-volume ratios differ
from those of the equal bipartitions used for even $N$, we present the
odd-size results separately rather than joining the two parity sectors
in a single finite-size plot.

All Hamiltonian parameters, initial states, averaging windows, and
quench-time optimization procedures are otherwise identical to those
used in the main text.

\subsection{Information hiding under SFIM evolution}
\begin{figure}[h]
    \centering
    \includegraphics[width=\linewidth]
    {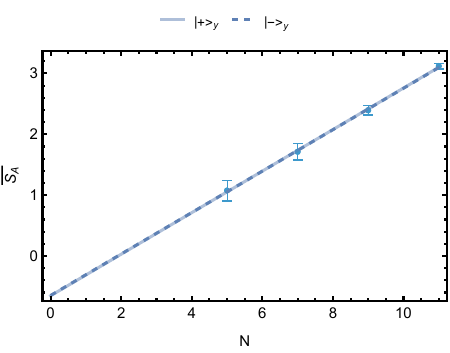}
    \includegraphics[width=\linewidth]
    {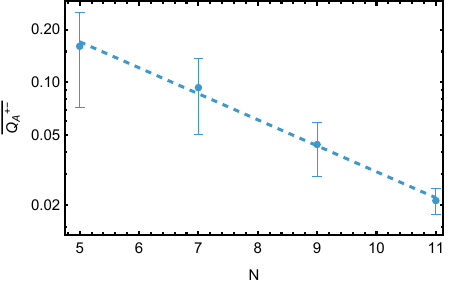}
    \caption{
    Information hiding under SFIM evolution for the odd system sizes
    $N=5,7,9,11$, with
    $N_A=\lfloor N/2\rfloor$.
    \textit{(top)} Half-chain von Neumann entropy for the two initial
    states.
    \textit{(bottom)} Hilbert--Schmidt distance between their reduced density
    matrices on subsystem $A$.
    The entropy growth and suppression of reduced-state distinguishability
    are qualitatively consistent with the even-size behavior presented
    in the main text.}
    \label{fig:odd-entropy-hs}
\end{figure}
Figure~\ref{fig:odd-entropy-hs} shows the half-chain entanglement entropy
and reduced-state Hilbert--Schmidt distance for the odd system sizes.
As for the even chains, the SFIM evolution rapidly generates
entanglement across the subsystem boundary while suppressing the
distinguishability of the two reduced states within the accessible
subsystem.

The entropy increases rapidly from its initially vanishing value,
demonstrating the generation of substantial bipartite entanglement.
At the same time, the Hilbert--Schmidt distance between the two
half-chain reduced states becomes strongly suppressed. Thus, the
initial-state dependence becomes difficult to access within the chosen
subsystem even though the global states remain orthogonal and the
evolution is fully unitary.

The odd-size results therefore support the same physical picture as
the even-size calculations: the SFIM dynamics delocalizes the
initial-state information into many-body correlations that are poorly
resolved by the reduced state and by the selected local measurement
channels.

\subsection{Standardized readout separation}

We next evaluate the Hamiltonian-switching readout using the same optimized quench time and metrics defined in the main text.

\begin{figure}[h]
    \centering
    \includegraphics[width=\linewidth]{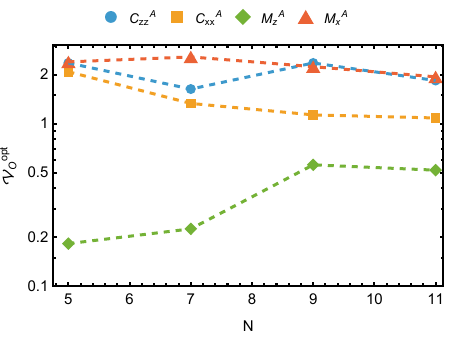}
    \includegraphics[width=\linewidth]{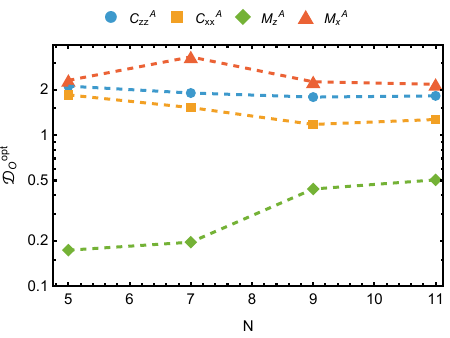}
    \caption{
    Standardized Hamiltonian-switching readout for the odd system sizes
    $N=5,7,9,11$.
    \textit{(top)} SFIM-normalized readout visibility
    $\mathcal{V}_O^{\rm opt}$.
    \textit{(bottom)} TFIM-normalized dynamical-band separation
    $\mathcal{D}_O^{\rm opt}$.
    The observables are $M_x^A$, $M_z^A$, $C_{xx}^A$, and $C_{zz}^A$. Dashed lines are guides to the eye.}
    \label{fig:odd-standardized-readout}
\end{figure}

The odd-size results reproduce the principal channel hierarchy found
for even systems. For $M_x^A$, the SFIM-normalized visibility remains
between approximately $1.94$ and $2.58$, while the TFIM-normalized
dynamical-band separation remains between approximately $2.17$ and
$3.29$. The $M_x^A$ readout therefore remains clearly resolved relative
to both the pre-quench and post-quench dynamical fluctuation scales
throughout the accessible odd-size range.

The $C_{zz}^A$ channel is similarly robust. Its SFIM-normalized
visibility lies between approximately $1.63$ and $2.36$, and its
TFIM-normalized separation lies between approximately $1.78$ and
$2.11$. Thus, the two strongest channels identified for even sizes,
$M_x^A$ and $C_{zz}^A$, also retain standardized separations
substantially above unity for all odd sizes considered.

For $C_{xx}^A$, both standardized measures remain above unity over the
accessible odd sizes, although the separation is generally smaller and
decreases from its value at $N=5$. By contrast, the $M_z^A$ channel
remains below one for both normalizations and is again the weakest
readout channel.

No simple monotonic finite-size law is apparent within the four
available odd sizes. Nevertheless, the standardized separation does
not exhibit a rapid collapse in the strongest measurement channels.
The odd-size calculations therefore support the conclusion drawn from
the even systems: although the absolute amplitudes of the
subsystem-averaged observables decrease with increasing system size,
the Hamiltonian switch continues to generate mean-signal separations
that are comparable to or larger than the relevant intrinsic dynamical
widths.

Taken together, the even- and odd-size results show that the observed readout enhancement is not tied to a particular parity or to the exact equal bipartition available for even chains. Across all observables considered, neither $\mathcal{V}_O^{\rm opt}$ nor $\mathcal{D}_O^{\rm opt}$ exhibits a clear monotonic decrease with increasing system size over the accessible range, although the magnitude of the standardized separation remains strongly observable dependent. The results therefore provide no evidence of a systematic finite-size collapse over $N=4$--$12$, while not establishing the asymptotic behavior at larger sizes. The central phenomenon, suppression of observable sensitivity under SFIM evolution followed by its enhancement under TFIM readout, is therefore common to both parity sectors.

\bibliographystyle{apsrev4-1}
\bibliography{Reference}

\end{document}